# High-fidelity conformation of graphene to $SiO_2$ topographic features

W.G. Cullen*,[1,2] M. Yamamoto,[1,2] K.M. Burson,[1,2] J.H. Chen,[1,2] C. Jang,[2] L. Li,[2] M.S. Fuhrer,[1,2] E.D. Williams [1,2]

[1] Materials Research Science and Engineering Center, Department of Physics, University of Maryland, College Park, Maryland 20742 USA
[2] Center for Nanophysics and Advanced Materials, Department of Physics, University of Maryland, College Park, Maryland 20742 USA

## Abstract

Strain engineering of graphene through interaction with a patterned substrate offers the possibility of tailoring its electronic properties, but will require detailed understanding of how graphene's morphology is determined by the underlying substrate. However, previous experimental reports have drawn conflicting conclusions about the structure of graphene on $SiO_2$. Here we show that high-resolution non-contact atomic force microscopy of $SiO_2$ reveals roughness at the few-nm length scale unresolved in previous measurements, and scanning tunneling microscopy of graphene on $SiO_2$ shows it to be slightly smoother than the supporting $SiO_2$ substrate. Quantitative analysis of the competition between bending rigidity of the graphene and adhesion to the substrate explains the observed roughness of monolayer graphene on $SiO_2$ as extrinsic, and provides a natural, intuitive description in terms of highly conformal adhesion. The analysis indicates that graphene adopts the conformation of the underlying substrate down to the smallest features with nearly 99% fidelity.

Graphene's morphology is expected to have a significant impact on its electronic properties[1, 2]. Local strain and curvature can introduce effective gauge fields, giving rise to carrier scattering and suppressing weak localization[3]. Engineered strained structures can realize artifical magnetic fields,with the possibility of a quantum Hall state in zero external field[4, 5]. It is therefore important to understand how graphene's morphology is determined by energetics. It has been suggested that free-standing graphene spontaneoulsy adopts a static corrugated morphology [6]. Other researchers have found this explanation unlikely, and have suggested an extrinsic origin for the corrugations observed in free standing graphene [7]. Regardless, when graphene is placed on a substrate, there is a strong expectation that graphene-substrate adhesion dominates the graphene morphology. This is qualitatively consistent with the observations of flat graphene on atomically-flat mica [8], and rough graphene on amorphous $SiO_2$ [9, 10]. However, the understanding of graphene's morphology on a rough substrate is far from settled, and the existing experimental reports [9, 10, 11] are at odds with each other. Here we show that the amorphous $SiO_2$ substrate is in fact much rougher than previously thought, and graphene supported on $SiO_2$ is slightly smoother than the underlying substrate. Graphene adopts the conformation of the underlying substrate down to the smallest features, consistent with the energetics of adhesion and bending, which indicate that graphene reproduces the substrate topography with nearly 99% fidelity.

In order for graphene to adopt a structure more corrugated than the underlying substrate, it must pay energy costs against both curvature and adhesion. Specifically, the adhesion energy $\gamma$ of graphene on $SiO_2$ has been estimated from carbon nanotube experiments [12, 13] and self-tensioning of suspended graphene resonators [14] to be $\approx 0.625$ eV/nm$^2$, and has been calculated for graphite/$SiO_2$ to be 0.5 eV/nm$^2$ [15]. The energy cost of bending graphene to form ripples is determined by distortion of the C-C bonds with the bending-induced loss of planarity, and the related strain (bond-length changes).

Direct calculations of the energy cost of bending graphene sheets [16, 17, 18] yield the uniaxial bending energy/area $E$ for a curvature $\kappa$, $E = 1/2C\kappa^2$, with $C = 0.85$ eV. Equating the adhesion energy $\gamma$ to the bending energy then allows a straightforward estimate of the curvature that can be maintained before the graphene ``pops free'' from the oxide substrate, yielding $1/\kappa = R \geq 0.9$ nm (or for symmetric biaxial bending, where $\kappa_x = \kappa_y, R_x \geq 1.3$ nm).

This simple analysis suggesting that graphene will adhere to the rough morphology of the $SiO_2$ down to the limit of structural features with a radius of curvature on the order of $R_{min} \approx 1$ nm will be expanded more quantitatively below, and the basic insight will be shown to hold. Even if the adhesion were an order of magnitude weaker, this conclusion would still hold to feature sizes of about 3 nm.

However, two earlier experiments have been interpreted in terms of an additional thermodynamic driving force which would cause graphene to adopt a non-flat conformation independent of substrate morphology [6, 11]. Transmission electron microscopy (TEM) of suspended graphene has shown structural tilts consistent with out-of-plane corrugations of $\approx$ 1 nm over a spatial extent of 10-25 nm [6]. A recent comparison of $SiO_2$ roughness measured at low resolution (using conventional AFM) with $SiO_2$-supported graphene measured with high resolution (STM) has shown a significantly larger measured roughness for the supported graphene [11]. This result was interpreted as indicating an intrinsic tendency toward graphene corrugation, possibly of the same origin as the corrugation observed in TEM. Here we will use high resolution UHV measurement for both $SiO_2$ and $SiO_2$-supported graphene to show that in fact the $SiO_2$ surface is rougher than previously known at the smaller lengths scales not accessed in lower resolution measurements. When both the graphene and the supporting substrate are measured with high resolution, the structure of the supported graphene closely matches that of the $SiO_2$ over the entire range of length scales, indicating that the graphene roughness observed is an extrinsic effect due to the $SiO_2$ substrate, and any intrinsic tendency toward corrugation of the graphene is overwhelmed by substrate adhesion.

Figure 1 compares scanned probe images of monolayer graphene on $SiO_2$ (STM) with bare $SiO_2$ (high resolution NC-AFM). The measured rms roughnesses are 0.35 nm and 0.37 nm, respectively, indicating that graphene is slightly smoother than the $SiO_2$. Further insight into the structure of graphene on $SiO_2$ may be gained by examining the Fourier spectra of the height data. Figure 2 shows the Fourier spectra for the images in Figures 1(a) and 1(b). The increased corrugation of the high resolution measurement of the oxide surface (Fig. 1(b)) is evident in the slightly increased amplitude of the Fourier spectrum (squares in Fig. 2) as compared to the graphene surface (Fig. 1(a) and triangles in Fig. 2). Also shown for comparison is the Fourier spectrum of a low-resolution measurement of the oxide surface (similar to that reported in Ref. [9]) which preserves the very long-wavelength structure (wavenumber $<$ 0.01 $nm^{-1}$) but clearly misses the structure which is seen by STM. The slightly decreased corrugation of graphene relative to the oxide surface below is expected based on a competition between adhesion energy and elastic curvature of the graphene sheet; this competition is discussed quantitatively below.

Understanding the length-scale dependence of the graphene morphology, and the slight decrease in graphene roughness compared with the substrate, requires a more sophisticated analysis. The approach needed can be drawn from analyses developed for membranes. In the absence of applied tension, the Hamiltonian is [19]:

$$H = \int \frac{C}{2}(\nabla^2 h(\mathbf{r}))^2 + V[h(\mathbf{r}) - z_s(\mathbf{r})]d^2 r. \qquad (1)$$

Here, $C$ is the elastic modulus (bending rigidity), $h(\mathbf{r})$ is the local height of the membrane and $z_s(\mathbf{r})$ is the local substrate height, both referenced to a flat reference plane ($\mathbf{r}$ represents spatial position in the 2-D reference plane). We make the simplifying assumption that the adhesion potential can be written as a function of the local relative height between substrate and overlayer.

We see that the energy is set by a combination of potential energy terms, which describe the balance between elastic energy in the graphene sheet and adhesion to the substrate. It is significant that the substrate adhesion term is a function of the local height difference between overlayer and substrate, whereas the elastic terms are directly obtained as functions only of the overlayer topography $h(\mathbf{r})$.

To address the first half of the energy balance, we compute the bending energy for the observed topography, hereafter referred to as curvature energy $E_C$. The statistical distributions of curvature for graphene and $SiO_2$ are shown in Figure 3. The curvature energy (per unit area) of the graphene overlayer is calculated as

$$E_C = \frac{C}{2}\left\{\frac{1}{A}\int (\nabla^2 h(\mathbf{r}))^2 d^2 r\right\}, \tag{2}$$

where $A$ is the area of the integration domain. The quantity in brackets evaluates to 0.078 $\mathrm{nm}^{-2}$ for the graphene topography corresponding to the image in Figure 1(a). Assuming that $C$ is bounded within 0.8 - 1.4 eV, we obtain 0.031 - 0.055 $\mathrm{eV/nm}^2$. Now, we may calculate the energy cost of *perfectly* following the substrate by making the substitution $h(\mathbf{r}) \to z_s(\mathbf{r})$ in Equation 2 above. For the $SiO_2$ topography shown in Figure 1(b), we obtain 0.092 - 0.161 $\mathrm{eV/nm}^2$ using the same bounds on $C$ (the quantity in brackets evaluates to 0.23 $\mathrm{nm}^{-2}$). Note that these values are obtained independently of any assumption about the adhesion energy, but their *difference* (avg = 0.084 $\mathrm{eV/nm}^2$) provides a value for the cost of curvature against the adhesion potential; the importance of evaluating this quantity will be shown below.

We now address the second half of the energy balance, which corresponds to the adhesion energy term. For analytic simplicity, we will work within the harmonic approximation for the adhesion potential. The calculation in Ref. [15] allows estimation of the harmonic coefficient $v = \partial^2 V/\partial z^2 |_{z=h_0}$ as $\approx 56$ $\mathrm{eV/nm}^4$ (see Supplement). Here, $h_0$ is the distance of the adhesion potential minimum from the substrate surface. The energy cost (per unit area) of deviating from the minimum in the adhesion potential is given by

$$\delta E_A = \frac{v}{2}\left\{\frac{1}{A}\int [h(\mathbf{r}) - z_s(\mathbf{r})]^2 d^2 r\right\}. \tag{3}$$

We define $\Delta(\mathbf{r}) = h(\mathbf{r}) - z_s(\mathbf{r})$, and see that the quantity in brackets is equivalent to the variance of $\Delta(\mathbf{r})$, given in terms of $h(\mathbf{r})$ and $z_s(\mathbf{r})$ as

$$\sigma_{\Delta}^{2}=\sigma_{h}^{2}+\sigma_{z_s}^{2}-2[<hz_s>-<h><z_s>] \quad (4)$$

where the final term may be removed by setting either $<h>$ or $<z_s>=0$. This expression makes clear that the variance in $\Delta(\mathbf{r})$ depends crucially on the degree of correlation between $h$ and $z_s$. While this correlation is not directly measurable using SPM, it is very instructive to consider its two limits.

In the limit of perfect adhesion (complete correlation), the variances $\sigma_h^2$ and $\sigma_{z_s}^2$ are cancelled by the correlation term $2<hz_s>$, and the variance $\sigma_\Delta^2$ vanishes ($\Delta(\mathbf{r})$ becomes a constant, $h_0$). One obtains the full adhesion energy (the complete depth of the potential well) in this case. However, in the uncorrelated limit ($<hz_s>=0$), it is apparent that the variance is very large because there is no cancellation.

The harmonic approximation is only valid for small excursions from the minimum in the potential well, and over-estimates the cost of large excursions away from the substrate. Taking an extreme limit of de-adhesion $\frac{1}{2}v\sigma_\Delta^2=0.6$ eV/nm$^2$, we see that this occurs for $\sigma_\Delta$ = 0.146 nm, equivalent to an amplitude of 0.207 nm. Thus, graphene with mean amplitude ~ 0.21 nm, uncorrelated with the $SiO_2$ substrate, would be essentially de-adhered from the substrate. This highlights the key physical concept that significant deviation from the substrate topography is quite costly, even for relatively weak interactions. Graphene which is highly uncorrelated with the underlying substrate would thus be adhered extremely weakly, which is very difficult to reconcile with the known adhesion properties of graphene and carbon nanotubes on $SiO_2$ [12-14].

In contrast, we now show that the energy balance is satisfied naturally by the highly conformal adhesion proposed here. The previous analysis of curvature energy gives the estimate 0.084 eV/nm$^2$ for the cost of curvature against the adhesion potential. This implies $\sigma_\Delta^2=0.003$ nm$^2$ according to Equation 4 (using $v=56$ eV/nm$^4$). A small positive number for $\sigma_\Delta^2$ is consistent with high (positive) correlation between $h$ and $z_s$. We obtain $<hz_s>/\sigma_h\sigma_{z_s}=0.99$, indicating an extremely high degree of correlation between graphene and the underlying substrate. The degree of correlation is likely overestimated due to the harmonic approximation for the adhesion potential. Although we cannot image the $SiO_2$ directly underneath the graphene, our analysis suggests that any topographic feature of the graphene must be the result of a corresponding substrate feature below.

We utilized the harmonic approximation as an analytical convenience for quantifying our topographic measurements. However, the description of highly conformal adhesion does not rely on this approximation, and we now discuss our results in the context of recent theories. For an adhesion energy near 0.5 - 0.6 eV/nm$^2$ and bending rigidity 1.4 - 1.5 eV, these unambiguously predict highly conformal adhesion [20, 21, 22]. In [21], the graphene-$SiO_2$ adhesion potential is described analytically by a Lennard-Jones pair potential, while in [20] a similar pair potential is used but with Monte Carlo integration over substrate atoms. Both [20] and [21] are parameterized in terms of the ratio $A_s/\lambda$ with the substrate modeled as a single-frequency sinusoidal corrugation with amplitude $A_s$ and wavelength $\lambda$. To compare with the experimental values, the $SiO_2$ topography exhibits power-law scaling with a correlation length of approximately 10 nm; if we associate the full rms roughness 0.37 nm with the wavelength 10 nm, we obtain $A_s/\lambda$ about 1/20. The adhesion transitions predicted in [20] and [21] only occur in the limit of much larger $A_s/\lambda$ or much weaker adhesion. Both [20] and [21] predict high conformation, with the ratio $A_g/A_s > 0.9$, where $A_g$ is the sinusoidal amplitude of graphene (g). The amplitude ratio determined from the measured rms roughnesses of graphene and $SiO_2$ is 0.95.

Conformal graphene adhesion is further predicted by an analysis which also includes the effects of periodic structures and accomodates nonlinear aspects of the membrane behavior [22]. There, it is shown that for a periodic sinusoidal profile, de-adhesion will occur in a series of transitions where first the membrane breaks loose from every other trough, then every two out of three troughs, and so on. In the zero-tension limit, these transitions are governed solely by the dimensionless parameter $\alpha$, where $\alpha = (\kappa_{eq}/\kappa_s)^{\frac{1}{2}}$. Here, $\kappa_{eq} = (2\gamma/C)^{\frac{1}{2}}$ with adhesion energy $\gamma$ and bending rigidity $C$ as above. $\kappa_s$ is the geometrical curvature of the substrate. A perfectly conforming ground state is predicted for $\alpha \geq 0.86$. Taking conservative estimates $\gamma = 0.5$ eV/nm$^2$ and $C = 1.4$ eV, the transition from perfect conformation occurs at substrate curvature $\kappa_s = 1.14$ nm$^{-1}$. Figure 3 shows a histogram of surface curvature obtained from high-resolution NC-AFM measurement of $SiO_2$. From the histogram it is apparent that curvatures exceeding 1.0 nm$^{-1}$ represent less than 0.1 percent of the $SiO_2$ surface topography.

Our preceding arguments have demonstrated that highly conformal adhesion to the $SiO_2$ substrate accounts for the observed graphene topography. This is primarily because the curvature energy scale set by the corrugation of $SiO_2$ is modest in comparison to that of the adhesion potential. With regard to the question of ``intrinsic'' rippling of graphene on $SiO_2$, we argue that this is physically unrealistic due to the overwhelming energy cost of deviating from the local minimum in $V(z)$. A recent calculation for isolated graphene finds the statically rippled structure energetically favorable by 0.0005 eV/nm$^2$ [23], miniscule in comparison to the energy scale set by adhesion. Further, we are unaware of any theory which predicts an *intrinsic* rippling of graphene when supported on a substrate, as additional energy would be required to offset the cost against bending energy and adhesion. A buckling instability does exist in the presence of external compressive strain, however [21].

Here we show that the corrugation observed for graphene on $SiO_2$ using STM is due to a higher corrugation of $SiO_2$, unresolved in previous AFM measurements. This presents a natural, intuitive description of exfoliated graphene topography based on established membrane physics. Our measurements agree with predicitions of three different theoretical models which use different parameterizations of the adhesion potential [20, 21, 22]. Careful consideration of the energy balance between bending rigidity and substrate adhesion shows that highly conformal adhesion gives a consistent description, whereas ``intrinsic'' rippling of graphene severely violates this balance. This interpretation is fully consistent with recent measurements of graphene exfoliated onto mica, which were found to be flat within 25 pm [8]. Even though our measurements of $SiO_2$ reveal higher corrugation than previously measured, the roughness is insufficient to induce the interesting structural transitions predicted by recent theories. Exploration of these issues, either by tailored substrates or chemical modification of graphene elasticity will provide interesting experiments for future work.

## Methods

The graphene samples were prepared by mechanical exfoliation of natural crystalline graphite on 300nm $SiO_2$ over doped silicon. Graphene monolayers were identified with an optical microscope and confirmed by Raman spectroscopy. The Raman spectra did not exhibit a d-peak, indicating that the observed area was defect free. To prevent photoresist contamination, Au/Ti electrodes were deposited through a shadow mask. The sample was cleaned by sonication in acetone followed by isopropyl alcohol. Measurements of $SiO_2$ were performed on a separate sample (from the same wafer batch) with the same solvent cleaning procedure. SPM measurements were carried out in an ultra-high vacuum measurement system which combines STM and NC-AFM (JEOL 4500A). STM was performed using a tip etched from Pt-Ir wire. High-resolution NC-AFM measurements of graphene and bare $SiO_2$ were performed using an AFM cantilever with super-sharp silicon tip (Veeco) having nominal radius of curvature 2 nm (upper limit 5 nm). These cantilevers have nominal force constant and resonant frequency 40 N/m and 300 kHz, respectively. For the graphene sample, measurements were performed with the contact and underlying Si substrate shorted together, and uniformly biased with respect to the SPM tip. SPM images are analyzed in raw form with only background subtraction. Further details on experimental protocol and data processing are discussed in the Supplement.

## Acknowledgements

This work has been supported by the University of Maryland NSF-MRSEC under grant no. DMR 05-20471, with supplemental funding from NRI. MRSEC Shared Experimental Facilities were used in this work. Additional infrastructure support provided by the UMD NanoCenter and CNAM. We would like to thank Nacional de Grafite for providing samples of natural graphite.

## Author Contributions

W.G.C., M.Y., K.M.B. and J.H.C. collected the data. W.G.C., M.Y., K.M.B., and E.D.W. analyzed the data. J.H.C., C.J. and L.L. prepared exfoliated graphene samples for the study, with assistance from M.Y. and K.M.B. W.G.C., M.S.F. and E.D.W. prepared the manuscript. W.G.C. directed the data collection and analysis, and E.D.W. directed the overall study. All authors contributed to the scientific process and editing of the manuscript.

## Additional Information

The authors declare no competing financial interests. Supplementary information accompanies this paper. Correspondence and requests for materials should be addressed to W.G.C. (wcullen@physics.umd.edu).

## Figures

(a)

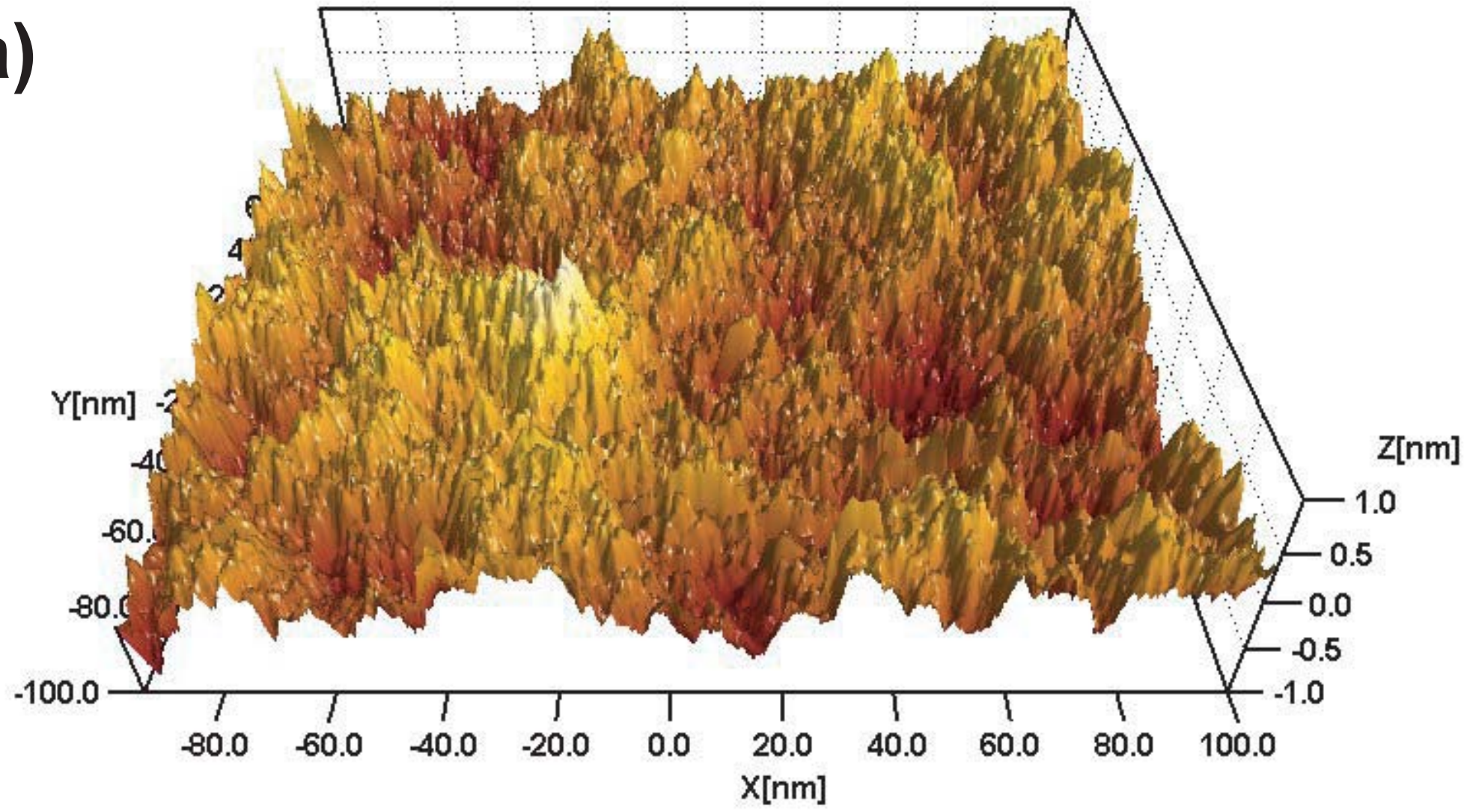

(b)

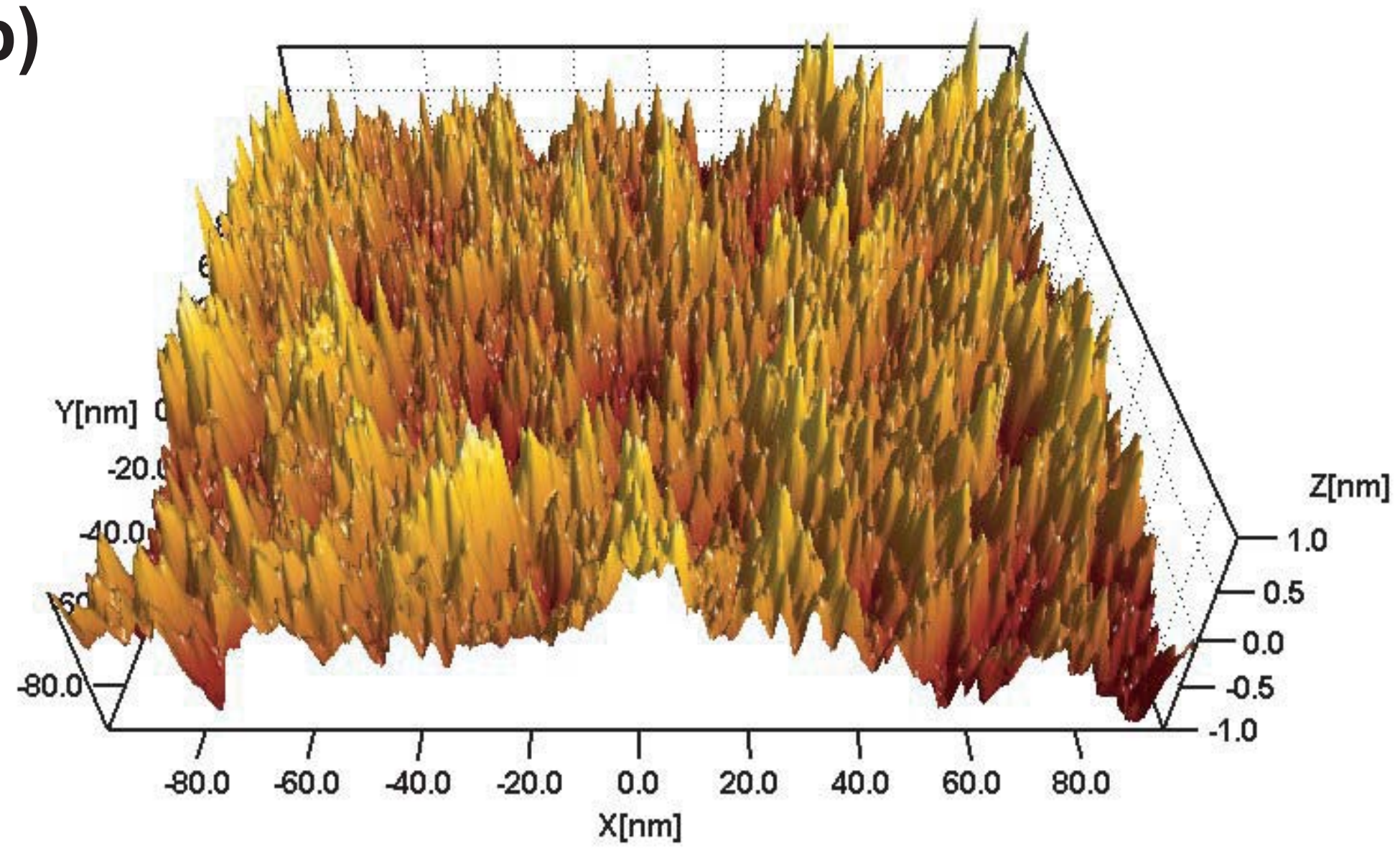

Figure 1: (color online) (a) STM of graphene monolayer ($195 \times 178$ nm, -305 mV, 41 pA) . (b) High-resolution NC-AFM of $SiO_2$ ($195 \times 178$ nm, $A = 5$ nm, $\Delta f = -20$ Hz).

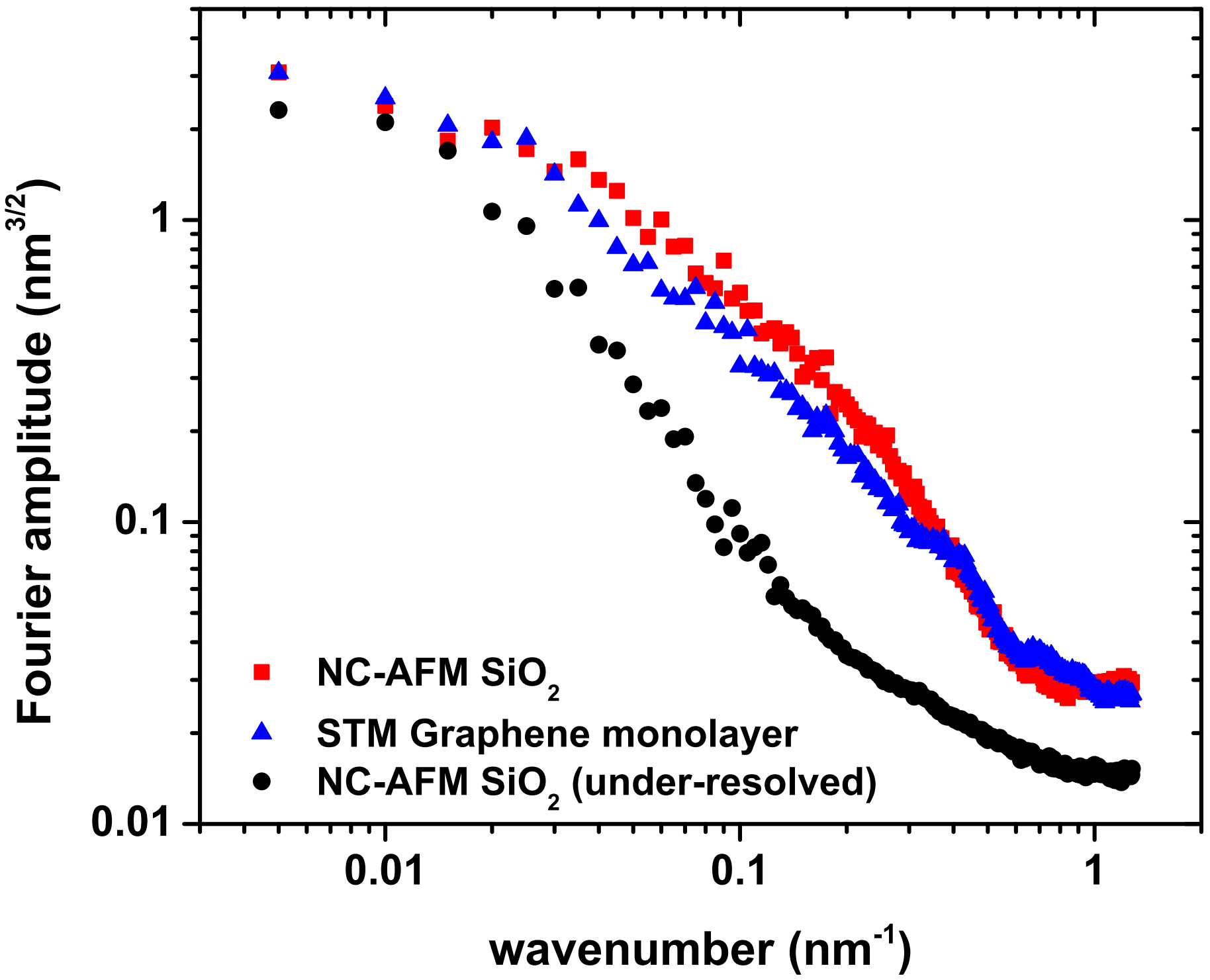

Figure 2: (color online) Fourier amplitude spectra, from top to bottom, of: $SiO_2$ NC-AFM ($\sigma_{rms} = 0.37$ nm), monolayer graphene STM ($\sigma_{rms} = 0.35$ nm), and under-resolved $SiO_2$ ($\sigma_{rms} = 0.22$ nm). Wavenumber is defined as wavelength$^{-1}$.

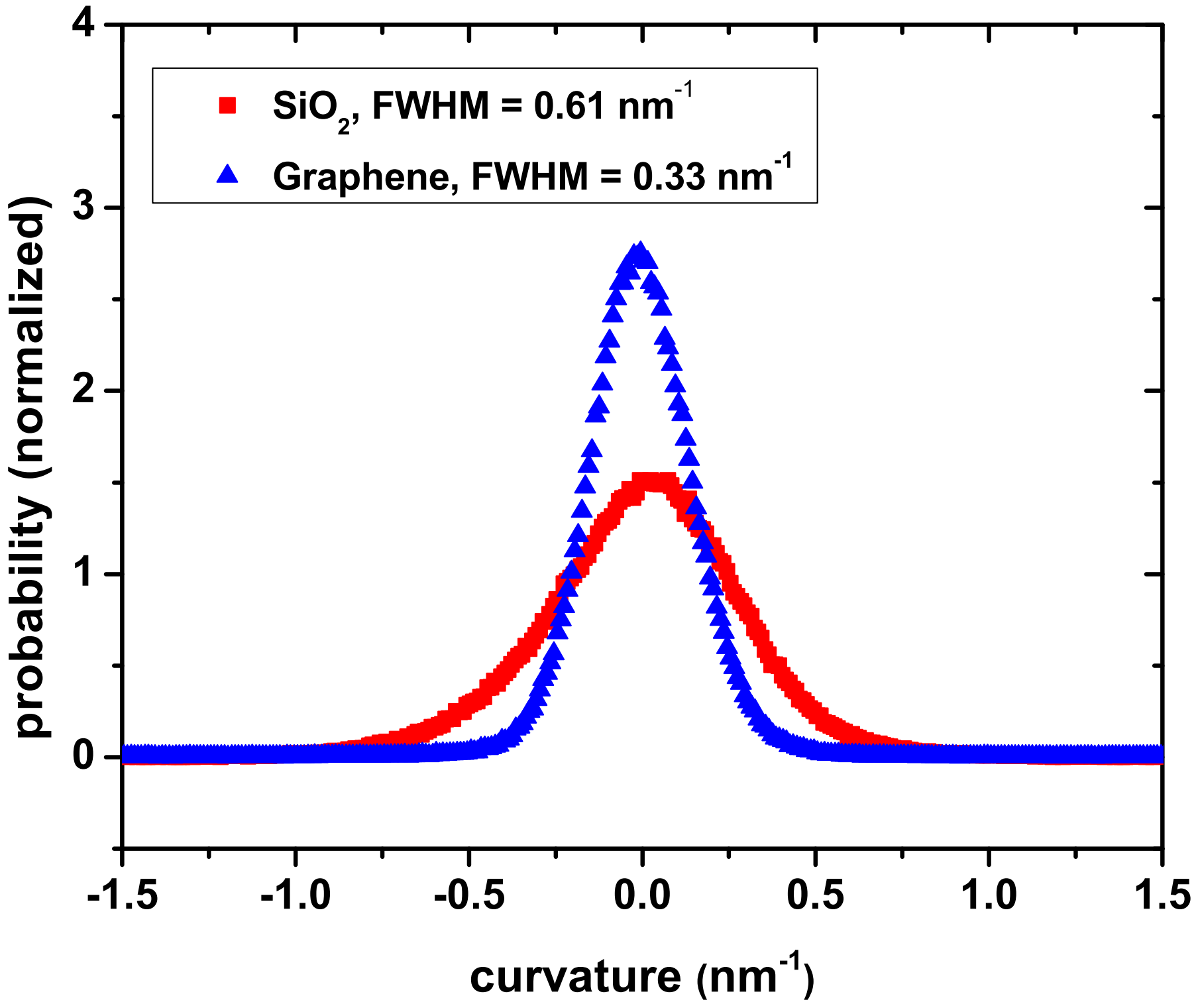

Figure 3: (color online) Curvature histograms, normalized to unit area, for graphene (narrower distribution) and $SiO_2$ (broader distribution).

# Supplementary information

Mon Jul 26 22:45:02 2010

This document contains supplementary information to accompany ``High-fidelity conformation of graphene to $SiO_2$ topographic features'' by Cullen et al.

## 1 Experimental methods

The graphene samples were prepared by mechanical exfoliation of natural crystalline graphite [1] on 300nm $SiO_2$ over doped silicon. Graphene monolayers were identified with an optical microscope and confirmed by Raman spectroscopy. The Raman spectra did not exhibit a d-peak, indicating that the observed area was defect free. To prevent photoresist contamination, Au/Ti electrodes were deposited through a shadow mask. The sample was cleaned by sonication in acetone followed by isopropyl alcohol. Measurements of $SiO_2$ were performed on a separate sample (from the same wafer batch) with the same solvent cleaning procedure.

SPM measurements were carried out in an ultra-high vacuum measurement system which combines STM and NC-AFM (JEOL 4500A)[2] equipped with Nanonis control electronics[3]. STM was performed using a tip etched from Pt-Ir wire. High-resolution NC-AFM measurements of graphene and bare $SiO_2$ were performed using an AFM cantilever with a super-sharp silicon tip (Veeco)[4] having nominal radius of curvature 2 nm (upper limit 5 nm). These cantilevers have nominal force constant and resonant frequency 40 N/m and 300 kHz, respectively. For the graphene sample, measurements were performed with the contact and underlying Si substrate shorted together, and uniformly biased with respect to the SPM tip. STM images are reported with tunnel current and bias voltage (applied to sample). NC-AFM images are reported with cantilever amplitude (nm) and frequency shift (Hz).

## 2 Data Analysis methods

SPM images were analyzed in raw form with only background subtraction. All data presented here were analyzed in one-dimensional fashion, along the fast-scan direction for SPM data acquisition. This is a well-established methodology which avoids certain artifacts in 2D analysis due to data anomalies in the slow-scan direction[5]. Thus, Fourier transforms presented here correspond to one-dimensional FFT's, averaged along the slow-scan direction. Where possible, data was additionally ensemble-averaged (over multiple images) to improve statistical certainty.

## 3 Comparison of oxide and graphene roughness

A central result of the present work is the observation that $SiO_2$ is in fact more corrugated than graphene over the relevant length scales $\approx 2-50$ nm. Here we provide a more complete summary of the experimental data which supports this conclusion. Figure 1 shows a series of Fourier amplitude spectra from images of graphene (STM) and $SiO_2$ (NC-AFM). Graphene samples 1 and 2 refer to separate exfoliated graphene samples with monolayer regions confirmed by Raman spectroscopy. Graphene samples were always heated to 130 C in UHV prior to imaging. $SiO_2$ sample 1 was heated at 130 C in UHV prior to imaging, then annealed at 500 C and subsequently imaged again. This is the designation ``before anneal'' and ``after anneal'' which appears in the graph legend for $SiO_2$ sample 1. In contrast, $SiO_2$ sample 2 was sonicated in acetone and isopropanol and quickly transferred into UHV with no further preparation. The dependence of topographic amplitude on annealing treatment for $SiO_2$ suggests that adsorbates may contribute to the measured $SiO_2$ roughness. Our UHV measurements of $SiO_2$ represent a ``lower bound'' for adsorbate effects in comparison to samples measured in ambient. Data comparisons at wavenumbers $>.5$ $nm^{-1}$ are difficult due to contributions from instrumental noise. However, we emphasize that in all measurements to date we have *not* been able to observe graphene rougher than the (fully resolved) $SiO_2$ for wavelengths ~ 10 nm, where intrinsic rippling is purported to dominate.

## 4 Membrane formalism

Consider the substrate + membrane system as indicated schematically in Figure 2. $h(\mathbf{r})$ is the local height of the membrane and $z_s(\mathbf{r})$ is the local substrate height, both referenced to a flat reference plane ($\mathbf{r}$ represents spatial position in the 2-D reference plane).

The general Hamiltonian describing membrane energetics is [6]:

$$H = \int \frac{C}{2}(\nabla^2 h(\mathbf{r}))^2 + \frac{\tau}{2}(\nabla h(\mathbf{r}))^2 + V[h(\mathbf{r}) - z_s(\mathbf{r})]d^2 r. \tag{1}$$

Here, $C$ is the elastic modulus (bending rigidity), and $\tau$ is the tension. In the absence of applied tension, the Hamiltonian reduces to

$$H = \int \frac{C}{2}(\nabla^2 h(\mathbf{r}))^2 + V[h(\mathbf{r}) - z_s(\mathbf{r})]d^2 r. \tag{2}$$

In order to simplify analytical calculations for membrane adhesion, a useful approximation (known as the Deryagin approximation) is to assume the potential is a simple function of the local relative height between substrate surface $z_s(r)$ and membrane $h(r)$. Within this approximation, and the harmonic approximation for potential defined below, the Hamiltonian becomes

$$H = \int \frac{C}{2}(\nabla^2 h(\mathbf{r}))^2 + \frac{v}{2}[h(\mathbf{r}) - z_s(\mathbf{r})]^2 d^2 r. \tag{3}$$

Here, $v$ is the harmonic potential coefficient (defined below).

## 5 Parameterization of adhesion energy

In this work we express all energies in units eV/nm$^2$, as appropriate for 2-D membranes. $V(z)$ describes the adhesion energy as a function of distance normal to the substrate surface. Qualitatively, a curve with a well-defined minimum is expected, as shown in Figure 2 of Ref. [7]. The adhesion energy $\gamma$ corresponds to the depth of the potential minimum. Carbon nanotube experiments [8, 9] and self-tensioning of suspended graphene resonators [10] yield estimates for $\gamma \approx 0.625$ eV/nm$^2$.

For analytic purposes, it is convenient to approximate the adhesion potential as a harmonic potential, which is parameterized by the harmonic coefficient $v$ (eV/nm$^4$). If $V(z)$ is the adhesion potential, we compute $v$ as

$$v = \frac{\partial^2 V}{\partial z^2}|_{z=h_0} . \tag{4}$$

where $h_0$ represents the height at which the potential is minimized. Since $V$ is expressed as energy/area, the harmonic coefficient $v$ has units Energy / (length)$^4$ and describes the cost of deviating from the minimum in the potential curve. One needs the potential curve in order to extract this coefficient, and Ref. [7] provides a computational result for graphite on silica (see Figure 2 in Ref. [7]). The potential is highly asymmetric in $z$, but can be approximated by a parabola for small excursions from the minimum. In Figure 3 we fit the points from Ref. [7] separately on both sides of the minimum. From these fits, the harmonic coefficient can be bounded between the values 76 eV per nm$^4$ and 30 eV per nm$^4$, depending on whether one takes the steep (repulsive) side of the potential (toward the substrate) or the shallower (attractive) side. The ``depth'' of the potential well is 0.5 eV/nm$^2$ (this corresponds to 80 mJ/m$^2$, the units expressed in Ref. [7]). Averaging these two estimates of $v$ yields the value 53 eV per nm$^4$.

Alternatively, one can arrive at this number by a more direct procedure, which emphasizes a useful scaling relation between these quantities. Following Aitken and Huang [11], one can begin with a Lennard-Jones potential written as

$$W_{LJ}(r) = -\frac{C_1}{r^6} + \frac{C_2}{r^{12}} \tag{5}$$

where it is assumed that the total energy is obtained pairwise over atoms in the substrate and graphene layer. For a flat monolayer and a flat substrate, the van der Waals potential becomes

$$U_{vdW}(z) = -\gamma[\frac{3}{2}(\frac{h_0}{z})^3 - \frac{1}{2}(\frac{h_0}{z})^9] \tag{6}$$

by integration over the 3-dimensional semi-infinite substrate. For the substrate plus overlayer geometry, $h_0$ is the equilibrium separation distance and $\gamma$ is the adhesion energy per unit area. By taking the second derivative with respect to $z$, one obtains the harmonic coefficient as

$$\frac{d^2U_{vdW}}{dz^2} = \frac{27\gamma}{h_0^2}[-\frac{2}{3}(\frac{h_0}{z})^5 + \frac{5}{3}(\frac{h_0}{z})^{11}] \tag{7}$$

We see that $\frac{d^2U_{vdW}}{dz^2}|_{z=h_0} = \frac{27\gamma}{h_0^2}$. In Ref. [7] they obtain $\gamma =$ 0.5 eV/nm $^2$ at equilibrium separation $h_0 = 0.492$ nm, which yields $v \approx 56$ eV/nm $^4$.

It should be mentioned that the harmonic approximation overestimates the energy cost of large excursions away from the substrate. Thus, it should be viewed as an approximation which is accurate for small fluctuations from the minimum in adhesion potential.

## Supplemental Figures

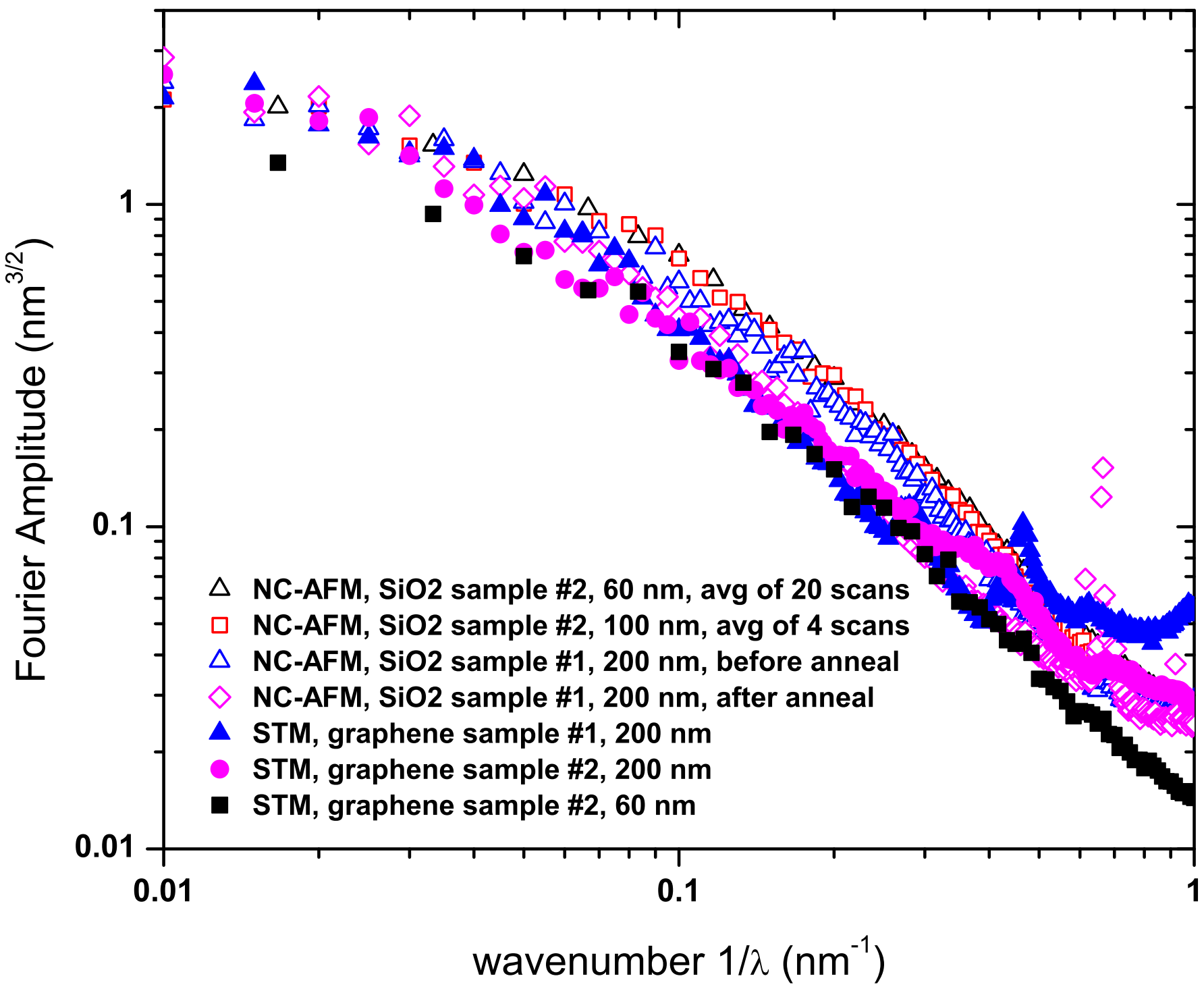

Figure 1: (color online) Extended comparison between $SiO_2$ and graphene. Solid symbols designate graphene STM measurement, open symbols designate $SiO_2$ NC-AFM measurement. Labels 60, 100, 200 nm indicate respective image size. See Supplementary text for discussion of $SiO_2$ annealing. The graphene STM image indicated by solid black squares is atomically resolved. All quantities here are analyzed from raw data, with no filtering (only background subtraction).

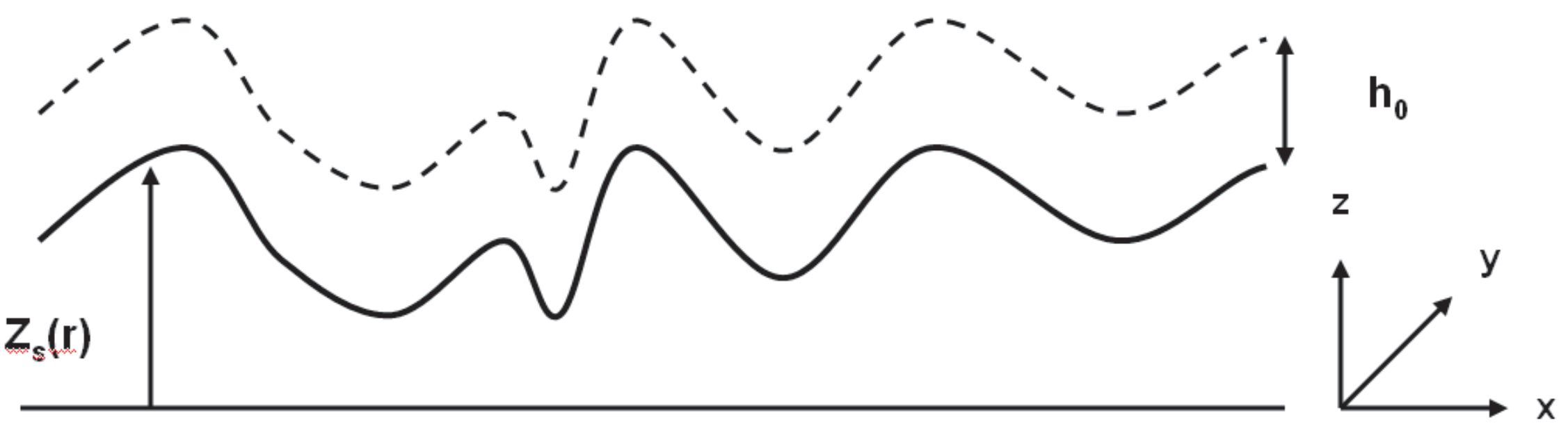

Figure 2: Membrane adhesion geometry. The substrate surface is at $z_s(\mathbf{r})$, and the local potential minimum for the membrane surface is at $z_s(\mathbf{r})+h_0$. The vector $\mathbf{r}$ is in the $x-y$ reference plane.

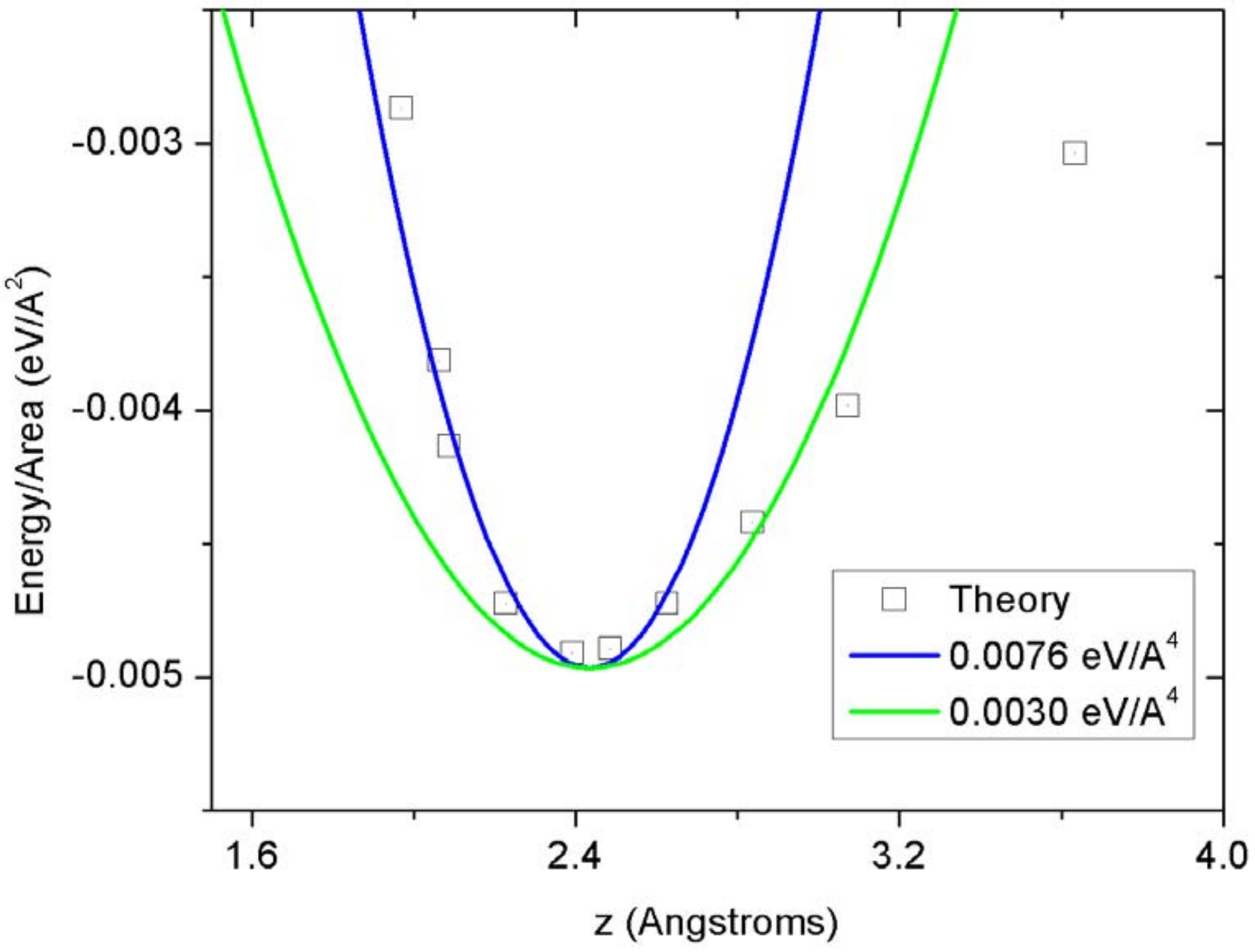

Figure 3: (Color online) Quadratic fits to adhesion energy curve from Ref. [7], in the vicinity of the potential minimum. Blue curve is fit to repulsive side of potential, green curve is fit to attractive side.